\documentclass[prl,aps,twocolumn,showkeys,superscriptaddress]{revtex4-2}
\usepackage{graphicx}
\usepackage{dcolumn}
\usepackage{bm}
\usepackage{amssymb}
\usepackage{amsmath}
\usepackage{booktabs}
\usepackage{hyperref}
\usepackage{multirow}
\usepackage{tikz}
\usetikzlibrary{3d, positioning, arrows.meta, shapes.geometric, plotmarks}

\usepackage{pgfplots}
\pgfplotsset{compat=1.18}

\begin{document}
\title{How elastic unitarity governs resonance peaks and residue phases}

\author{S.~Ceci}
\email{sasa.ceci@irb.hr}
\affiliation{Rudjer Bo\v{s}kovi\'{c} Institute, Bijeni\v{c}ka 54, HR-10000 Zagreb, Croatia}
\author{R.~Omerovi\'{c}}
\affiliation{University of Tuzla, Urfeta Vejzagi\'{c}a 4, 75000 Tuzla, Bosnia and Herzegovina}
\author{H.~Osmanovi\'{c}}
\affiliation{University of Tuzla, Urfeta Vejzagi\'{c}a 4, 75000 Tuzla, Bosnia and Herzegovina}
\author{M.~Uroi\'{c}}
\affiliation{Rudjer Bo\v{s}kovi\'{c} Institute, Bijeni\v{c}ka 54, HR-10000 Zagreb, Croatia}
\author{B.~Zauner}
\affiliation{Institute for Medical Research and Occupational Health, Ksaverska 2, HR-10000 Zagreb, Croatia}
\date{\today}

\begin{abstract}
Imposing elastic unitarity on resonant amplitudes yields a geometric rule connecting the reaction threshold, S-matrix pole, and Breit-Wigner peak. Validated across six orders of magnitude in energy, from $^5\text{He}$ to the Higgs boson, this rule predicts currently unknown residue phases: $-44(12)^\circ$ for $^5\mathrm{He}$, $-24(7)^\circ$ for $\Sigma(1385)^+$, and $-7(13)^\circ$ for $\Xi(1530)^0$. By inverting this formalism, we determine the $\Upsilon(4S)$ pole from its empirical peak to be $10575(1)-i\,8.3(13)$ MeV, with a $-52(6)^\circ$ phase.
\end{abstract}

\maketitle


\textit{Introduction.}---Resonances are ubiquitous across all energy scales in nuclear and particle physics, from nuclear systems (see e.g.,~\cite{Tilley2002, Bond1977}), among mesons and baryons, all the way up to gauge and Higgs bosons~\cite{PDG2026}. However, the physical meaning of key parameters like the pole residue phase remains elusive. The standard Breit-Wigner (BW) parametrization~\cite{BW}, often enhanced with energy-dependent widths~\cite{Jackson1964} or Flatté modifications~\cite{Flatte1976} to account for channel openings, remains the primary tool for extracting resonant properties from experimental cross sections. While directly observable, this empirical peak is not fundamental. A resonance is rigorously defined by its complex S-matrix pole, hidden on the second (nonphysical) Riemann sheet. Furthermore, the classical Breit-Wigner mass suffers from well-known gauge-invariance issues~\cite{Sirlin1991,Djukanovic2007}. This problem may, nevertheless, be resolved by redefining the BW mass not through the propagator, but via the zero-crossing of the full scattering amplitude's real part~\cite{Djukanovic2007}.

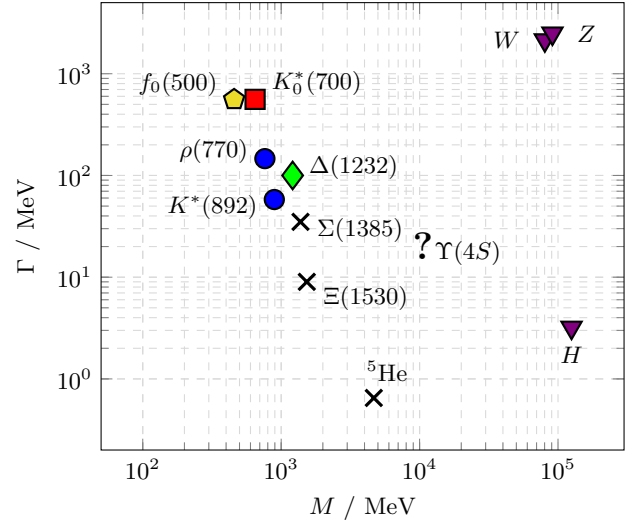
\begin{figure}[tbp]
\centering
\begin{tikzpicture}
\begin{axis}[
xmode=log,
ymode=log,
width=8.5cm,
height=7.5cm,
xmin=50, xmax=300000, 
ymin=0.2, ymax=5000,
xlabel={$M$ / MeV},
ylabel={$\Gamma$ / MeV},
grid=both,
grid style={dashed, gray!30},
title={$M-\Gamma$ diagram (no apparent correlation)},
title style={font=\bfseries, yshift=1ex},
xtick={100, 1000, 10000, 100000}, 
ytick={1, 10, 100, 1000},
axis background/.style={fill=white},
clip=false
]

\addplot[only marks, mark=x, mark options={draw=black, very thick, scale=2.2}] coordinates { (4667.8, 0.65) }; \node[above, xshift=5pt, yshift=3pt] at (4667.8, 0.65) {$^5\text{He}$};
\addplot[only marks, mark=triangle*, mark options={draw=black, thick, fill=violet, scale=2.2, rotate=180}] coordinates { (80341, 2140) (91162, 2494) (125250, 3.2) }; \node[left, xshift=-6pt] at (80341, 2140) {$W$}; \node[right, xshift=6pt] at (91162, 2494) {$Z$}; \node[below, yshift=-4pt] at (125250, 3.2) {$H$};
\addplot[only marks, mark=x, mark options={draw=black, very thick, scale=2.2}] coordinates { (1531.6, 9.0) }; \node[right, xshift=3pt, yshift=-6pt] at (1531.6, 9.0) {$\Xi(1530)$};
\addplot[only marks, mark=*, mark options={draw=black, thick, fill=blue, scale=1.8}] coordinates { (762.5, 146.4) (892, 58) }; \node[left, xshift=-3pt, yshift=3pt] at (762.5, 146.4) {$\rho(770)$}; \node[left, xshift=-3pt, yshift=-3pt] at (892, 58) {$K^*(892)$};
\addplot[only marks, mark=x, mark options={draw=black, very thick, scale=2.2}] coordinates { (1379, 35) }; \node[right, xshift=3pt, yshift=-3pt] at (1379, 35) {$\Sigma(1385)$};
\addplot[only marks, mark=diamond*, mark options={draw=black, thick, fill=green, scale=2.6}] coordinates { (1210, 100) }; \node[right, xshift=3pt, yshift=3pt] at (1210, 100) {$\Delta(1232)$};
\addplot[only marks, mark=pentagon*, mark options={draw=black, thick, fill=yellow!90!black, scale=2.0}] coordinates { (457, 558) }; \node[left, xshift=-3pt, yshift=6pt] at (457, 558) {$f_0(500)$};
\addplot[only marks, mark=square*, mark options={draw=black, thick, fill=red, scale=1.8}] coordinates { (648, 560) }; \node[right, xshift=3pt, yshift=6pt] at (648, 560) {$K_0^*(700)$};
\node[font=\bfseries\Large, text=black] at (10579, 20.5) {?}; \node[right, xshift=1pt, yshift=-3pt] at (10579, 20.5) {$\Upsilon(4S)$};
\end{axis}
\end{tikzpicture}
\caption{Resonant pole position diagram. Marker shapes and colors denote residue half-phase ranges ($\theta/2$): violet inverted triangles ($0^\circ$ to $-5^\circ$), blue circles ($-5^\circ$ to $-15^\circ$), green diamonds ($-15^\circ$ to $-45^\circ$), yellow pentagons ($-45^\circ$ to $-75^\circ$), orange stars ($-75^\circ$ to $-85^\circ$)---none of the resonances in this study with a known $\theta/2$ fall into this range---and red squares ($-85^\circ$ to $-90^\circ$). For consistency, for $^5\text{He}$ we use its relativistic mass rather than the conventional kinetic energy. Black crosses ($\times$) represent states with unknown residue phases. The heavy $\Upsilon(4S)$ state is denoted by a prominent question mark (?) because extracting its S-matrix pole directly from empirical Breit-Wigner parameters is a notoriously difficult problem, leaving it unresolved here.}
\label{fig:apparent_chaos}
\end{figure}

An intriguing open question is whether the pole position $E_p = M-i\,\Gamma/2$ and the pole residue $r=|r|e^{i\theta}$ are independent. Indeed, plotting resonances solely by their pole mass and width reveals no apparent correlation with their elastic residue phases (see Fig.~\ref{fig:apparent_chaos}). A similar puzzle surrounds the observable peaks: why is the peak sometimes near the pole mass ($\Xi(1530)$), sometimes shifted further away ($\Delta(1232)$), and sometimes completely absent ($f_0(500)$~\cite{GM}, $K_0^*(700)$~\cite{PR})? Such distortions make extracting S-matrix poles directly from empirical BW parameters notoriously difficult, leaving states like the heavy quarkonium $\Upsilon(4S)$ unresolved. And lastly, are resonance phenomena fundamentally different in nuclear physics (e.g., the $^5\mathrm{He}$ state in $n\alpha$ scattering) compared to elementary particles like the $W$, $Z$, and Higgs bosons?

In this Letter, we show that each of these questions has a clear answer for elastic resonances. We build our model on a phenomenological observation that a first-order Laurent expansion of the scattering amplitude near the pole naturally generates an effective zero near the reaction threshold~\cite{Ceci13}. This zero, together with the observable peak position, is then linked to the complex S-matrix pole via a semi-empirical formula~\cite{Ceci2017}, recently generalized into a broader mathematical framework~\cite{Ceci2026_PLB1, Ceci2026_PLB2}. By imposing S-matrix unitarity ($S^\dagger S=I$) on this form, we derive a relation connecting all key resonant parameters: the threshold, the pole and its residue, and the position and width of the experimental peak. We show this relation is valid for various resonances across six orders of magnitude in energy. By inverting this relation, we provide a simple tool to estimate pole parameters directly from empirical peak properties, and apply it to the $\Upsilon(4S)$ resonance.

\textit{Model.}---A resonance is unambiguously defined by a complex S-matrix pole ($S=I+2iT$). In practice, however, experimental analyses typically fit the observable cross sections using the Breit-Wigner mass $M_\mathrm{BW}$ and width $\Gamma_\mathrm{BW}$. While $M_\mathrm{BW} \approx M$ for isolated narrow resonances far from thresholds, it is known that non-resonant backgrounds can severely shift this observable peak, especially for broad, near-threshold states.

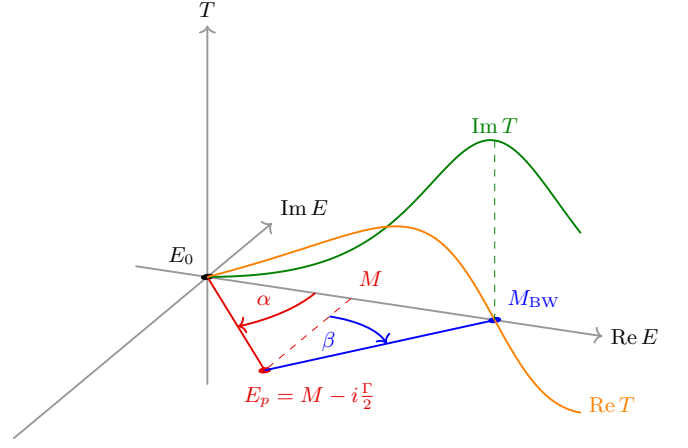
\begin{figure}[tbp]
\centering
\resizebox{\columnwidth}{!}{%
\begin{tikzpicture}[x={(1cm,-0.15cm)}, y={(0.6cm,0.5cm)}, z={(0cm,1cm)}, scale=1.05]
\draw[->, thick, gray!80] (1.0, 0, 0) -- (7.5, 0, 0) node[right, text=black] {$\mathrm{Re}\, E$};
\draw[->, thick, gray!80] (2, -4.5, 0) -- (2, 1.5, 0) node[above right, text=black] {$\mathrm{Im}\, E$};
\draw[->, thick, gray!80] (2, 0, -1.5) -- (2, 0, 3.5) node[above, text=black] {$T$};

\begin{scope}[canvas is xy plane at z=0]
\filldraw[black] (2, 0) circle (2pt) node[above left, xshift=-2pt, yshift=2pt] {$E_0$};
\filldraw[blue] (6, 0) circle (2pt) node[above right, xshift=2pt, yshift=2pt] {$M_\mathrm{BW}$};

\draw[dashed, red!90!black] (4, 0) node[above right, yshift=2pt, text=red!90!black] {$M$} -- (4, -2);

\filldraw[red!90!black] (4, -2) circle (2pt) node[below right, xshift=-12pt, yshift=-2pt, text=red!90!black] {$E_p = M - i\frac{\Gamma}{2}$};

\draw[thick, red!90!black] (2, 0) -- (4, -2);
\draw[thick, blue] (6, 0) -- (4, -2);

\draw[->, thick, red!90!black] (3.5, 0) arc (0:-45:1.5);
\node[red!90!black] at (3.0, -0.35) {$\alpha$};
\draw[->, thick, blue] (4, -0.5) arc (90:45:1.5);
\node[blue] at (4.35, -1.1) {$\beta$};
\end{scope}

\begin{scope}[canvas is xz plane at y=0]
\draw[thick, green!50!black, smooth, domain=2:7.2, samples=80] 
plot (\x, {2.5 * (1 - ((\x-6)/4)^2) / (1 + ((\x-6)/1.5)^2)});
\node[green!50!black, above] at (6, 2.5) {$\mathrm{Im}\, T$};

\draw[thick, orange, smooth, domain=2:7.2, samples=80] 
plot (\x, {-2.5 * (1 - ((\x-6)/4)^2) * ((\x-6)/1.5) / (1 + ((\x-6)/1.5)^2)});
\node[orange, right] at (7.2, -1.0) {$\mathrm{Re}\, T$};

\draw[dashed, green!50!black] (6, 2.5) -- (6, 0);
\end{scope}
\end{tikzpicture}%
}
\caption{Geometric representation of resonance parameters. The scattering amplitude $T$ starts from an effective zero at the threshold $E_0$, and its real part crosses zero at the Breit-Wigner mass $M_\mathrm{BW}$, which defines the observable resonance peak position. In the complex energy plane, the pole is located at $E_p = M-i\Gamma/2$, defining angles $\alpha$ and $\beta$. Elastic unitarity dictates the geometric symmetry $\beta = \alpha$.}
\label{fig:complex_plane}
\end{figure}

Our model rests on a simple premise: the scattering amplitude $T$ must possess a resonance pole, a reaction threshold zero, and a constant background $T_B$. The simplest form satisfying these conditions is given by $T(E) = (E - E_0)/(E - E_p)\,T_B$. This constant background must satisfy the same elastic unitarity condition as the full amplitude, $\mathrm{Im}\, T_B = |T_B|^2$ (due to asymptotic behavior $T(E\rightarrow\infty)=T_B$). The complex energy plane ($\mathrm{Im} \,E$ vs.~$\mathrm{Re} \,E$) and the scattering amplitude plane ($T$ vs.~$\mathrm{Re} \,E$) are illustrated in Fig.~\ref{fig:complex_plane}.

Defining the amplitude peak position $M_\mathrm{BW}$ as the zero-crossing of its real part ($\mathrm{Re}\, T(M_\mathrm{BW})=0$) naturally generates two geometric angles, $\alpha$ and $\beta$~\cite{Ceci2017}:
\begin{equation}
\tan \alpha = -\frac{\Gamma/2}{M - E_0}, \quad \tan \beta = -\frac{M_\mathrm{BW} - M}{\Gamma/2}.\label{eq:alphabeta}
\end{equation}
Imposing S-matrix unitarity on the scattering amplitude $T$ demands $\beta=\alpha$.

Since this model is an approximation based on the first-order Laurent expansion around the pole position, in practice we would expect $\alpha$ and $\beta$ to slightly disagree, and that $\beta$ is the more robust value. Therefore, the elastic residue phase that is generally given by $\theta \approx \alpha + \beta$~\cite{Ceci2017,Ceci2026_PLB1} is, for elastic resonances, closer to $\theta = 2\beta$, perfectly reproducing H\"ohler's result for adding a simple pole to a constant background with phase $\beta$~\cite{Hoehler1983}. 

By equating $\alpha$ and $\beta$ in Eq.~(\ref{eq:alphabeta}), we can predict the location of the peak (a parameter we name the unitary Breit-Wigner mass $\widetilde{M}_\mathrm{BW}$) purely from the pole and threshold parameters:
\begin{equation}
\widetilde{M}_\mathrm{BW} = M + \frac{(\Gamma/2)^2}{M - E_0}.\label{eq:MBWU}
\end{equation}

Manley previously introduced a parameter equivalent to $\beta$ to deduce pole coordinates from empirical BW parameters~\cite{Manley} for nucleon resonances. Lacking proper values, he extracted it from pole and BW parameters in concurrent analyses. We resolve this by sticking to elastic resonances, where $\beta = \alpha$.

Following Manley's route, the S-matrix pole ($\widetilde{M}, \widetilde{\Gamma}$) can be directly reconstructed from empirical parameters if $\beta$ is known:
\begin{equation}
\widetilde{M} = M_\mathrm{BW} + \frac{\Gamma_\mathrm{BW}}{2}\sin\beta\cos\beta, \quad \widetilde{\Gamma} = \Gamma_\mathrm{BW}\cos^2\beta. \label{eq:pole_reconstruction}
\end{equation}

Normally, this would be a serious caveat for us, since there is no general way to know $\beta$. However, for elastic resonances, by substituting the width from Eq.~(\ref{eq:pole_reconstruction}) into Eq.~(\ref{eq:alphabeta}), and enforcing the elastic condition $\beta \equiv \alpha$, we express $\beta$ solely in terms of empirical BW parameters:
\begin{equation}
\tan\beta \equiv\tan \alpha= -\frac{\Gamma_\mathrm{BW}/2}{M_\mathrm{BW} - E_0}.\label{eq:alphaBW}
\end{equation}
This allows us to extract pole coordinates and residue phases ($\theta=2\beta$) directly from elastic BW peaks, providing a practical tool for states like $\Upsilon(4S)$ where direct pole extractions are unavailable.

\begin{table*}[tbp]
\centering
\caption{All input parameters, confirmations, and predictions. The unitary peak $\widetilde{M}_\mathrm{BW}$ is given by Eq.~(\ref{eq:MBWU}). The parameter $\xi_E$ from Eq.~(\ref{eq:ksiE}) flags departures from elastic unitarity. \textbf{Part I:} Nuclear resonance and canonical elastic hadrons obey the geometric symmetry ($\xi_E \approx 0$). H\"ohler's rule ($\theta = 2\beta$) robustly predicts their unitary residue phases. \textbf{Part II:} Symmetry universally extends to heavy elementary particles. \textbf{Part IIIa:} For threshold-dominated states ($f_0(500)$, $K_0^*(700)$), PDG empirical parameters yield large unitarity departures ($\xi_E \gg 0$). \textbf{Part IIIb:} By discarding the classical unobservable $M_\mathrm{BW}$ and enforcing the exact elastic geometry ($\beta \equiv \alpha$), we calculate their unitary peaks ($\widetilde{M}_\mathrm{BW}$), predicting residue phases (as $2\alpha$) and pole coordinates. All energies are in MeV.}
\label{tab:main_results}
\begin{tabular*}{\textwidth}{@{\extracolsep{\fill}} lcccccccccc}
\toprule
\textbf{State} & $\boldsymbol{E_0}$ & $\boldsymbol{M}$ & $\boldsymbol{\Gamma}$ & $\boldsymbol{M_\mathrm{BW}}$ & $\boldsymbol{\widetilde{M}_\mathrm{BW}}$ & $\boldsymbol{-\alpha / {}^\circ}$ & $\boldsymbol{-\beta / {}^\circ}$ & $\boldsymbol{\xi_E}$ & $\boldsymbol{-2\beta / {}^\circ}$ & $\boldsymbol{-\theta_{Ref} / {}^\circ}$ \\
\midrule
\multicolumn{11}{c}{\textbf{PART I: NUCLEAR AND HADRON ELASTIC RESONANCES $\bm{(\xi_E\approx0)}$ }}
\\
\midrule
$^5$He & $0$\textsuperscript{a} & $0.80(2)$ & $0.65(5)$ & $0.93(3)$ & $\boldsymbol{0.93(3)}$ & $22(2)$ & $22(6)$ & $\bm{0.00(12)}$ & $\bm{44(12)}$ & ---\textsuperscript{b} \\
$\rho(770)^0$ & $279$ & $762.5(17)$ & $146.4(22)$ & $775.3(2)$ & $\boldsymbol{773.6(17)}$ & $8.6(1)$ & $10(1)$ & $\bm{0.02(2)}$ & $\bm{20(3)}$ & $19$\textsuperscript{c} \\
$K^*(892)^0$ & $633$ & $892(1)$ & $58(2)$ & $895.6(2)$ & $\boldsymbol{895.2(10)}$ & $6.4(2)$ & $7(2)$ & $\bm{0.01(4)}$ & $\bm{14(4)}$ & $15$\textsuperscript{c} \\
$\Delta(1232)$ & $1078$ & $1210(1)$ & $100(2)$ & $1232(2)$ & $\boldsymbol{1229(1)}$ & $20.7(4)$ & $24(2)$ & $\bm{0.06(5)}$ & $\bm{47(4)}$ & $46^{+2}_{-1}$\textsuperscript{d} \\
$\Sigma(1385)^+$ & $1255$ & $1379(1)$ & $35(3)$ & $1382.8(3)$ & $\boldsymbol{1381.5(11)}$ & $8.0(7)$ & $12(3)$ & $\bm{0.08(6)}$ & $\bm{24(7)}$ & ---\textsuperscript{b} \\
$\Xi(1530)^0$ & $1461$ & $1531.6(4)$ & $9.0(8)$ & $1531.8(3)$ & $\boldsymbol{1531.9(4)}$ & $3.7(3)$ & $3(7)$ & $\bm{0.0(1)}$ & $\bm{7(13)}$ & ---\textsuperscript{b} \\
\midrule
\multicolumn{11}{c}{\textbf{PART II: HEAVY ELEMENTARY PARTICLES 
$\bm{(\xi_E\approx 0)}$ 
}} \\
\midrule
$W$ boson & $\sim 0$ & $80341(8)$ & $2140(50)$ & $80362(8)$ & $\boldsymbol{80355(8)}$ & $0.76(2)$ & $1.15(6)$ & $\bm{0.007(1)}$ & $\bm{2.3(1)}$ & $2.29(5)$\textsuperscript{e} \\
$Z$ boson & $\sim 0$ & $91162(2)$ & $2494(2)$ & $91188(2)$ & $\boldsymbol{91179(2)}$ & $0.78(1)$ & $1.18(1)$ & $\bm{0.007(1)}$ & $\bm{2.36(1)}$ & $2.35(1)$\textsuperscript{e} \\
$H$ boson & $\sim 0$ & $125250(170)$ & $3.2(22)$ & $\bm{\equiv\,M}$ & $\boldsymbol{125250(170)}$ & $0.00(0)$ & $\bm{\equiv 0}$ & $\bm{0.00(0)}$ & $\bm{\equiv 0}$ & $0.00(0)$\textsuperscript{e} \\
\midrule
\multicolumn{11}{c}{\textbf{PART IIIa: ELASTIC RESONANCES WITH METRIC
$\bm{\xi_E \gg 0}$ OR UNDEFINED 
}} \\
\midrule
$f_0(500)$ & $279$ & $457(14)$ & $558(22)$ & $475(75)$\textsuperscript{f} & --- & $57(2)$ & $4(16)$ & $\bm{1.5(3)}$ & $\bm{7(31)}$ & $118(12)$\textsuperscript{c} \\
$K_0^*(700)$ & $633$ & $648(7)$ & $560(32)$ & $680(50)$\textsuperscript{f} & --- & $87(1)$ & $7(10)$ & $\bm{19(9)}$ & $\bm{13(20)}$ & $180(8)$\textsuperscript{c} \\
$\Upsilon(4S)$\textsuperscript{g} & $10558$ & --- & --- & $10579(1)$\textsuperscript{f} & --- & --- & --- & --- & --- & ---\textsuperscript{b} \\
\midrule
\multicolumn{11}{c}{\textbf{PART IIIb: PARAMETERS FIXED AFTER IMPOSING UNITARITY 
$\bm{(\beta \equiv \alpha\rightarrow\xi_E\equiv 0)}$ 
}} \\
\midrule
$f_0(500)$ & $279$ & $\boldsymbol{457(14)}$ & $\boldsymbol{558(22)}$ & --- & $\boldsymbol{894(40)}$ & $57(2)$ & $\boldsymbol{\equiv \alpha}$ & $\bm{\equiv 0}$ & $\boldsymbol{115(5)}$\textsuperscript{h} & $118(12)$\textsuperscript{c} \\
$K_0^*(700)$ & $633$ & $\boldsymbol{648(7)}$ & $\boldsymbol{560(32)}$ & --- & $\boldsymbol{\sim \infty}$ & $87(1)$ & $\boldsymbol{\equiv \alpha}$ & $\bm{\equiv 0}$ & $\boldsymbol{174(3)}$\textsuperscript{h} & $180(8)$\textsuperscript{c} \\
$\Upsilon(4S)$ & $10558$ & $\boldsymbol{10575(1)}$\textsuperscript{i} & $\boldsymbol{16.6(13)}$\textsuperscript{i} & --- & $\boldsymbol{10579(1)}$ & $26(3)$ & $\boldsymbol{\equiv \alpha}$ & $\bm{\equiv 0}$ & $\boldsymbol{52(6)}$\textsuperscript{h} & ---\textsuperscript{b} \\  
\bottomrule
\end{tabular*}

\smallskip
\footnotesize
\raggedright
\textsuperscript{a} Masses for $^5\text{He}$ are displayed relative to the $n-\alpha$ scattering threshold ($E_0 \equiv 0$) following standard nuclear convention. Its pole mass (i.e., its rest mass) in our convention would be $M \approx 4667.8$ MeV (with the threshold at $E_0 \approx 4667.0$ MeV). \\[2pt]
\textsuperscript{b} Theoretical prediction of the unitary residue phase (no experimental extraction available).\\[2pt]
\textsuperscript{c} Fundamental pole parameters ($M$, $\Gamma$) and complex residue phases extracted together from recent dispersive analyses~\cite{GM, PR}. The phases are translated to our $E$-plane convention via Eq.~(\ref{eq:convention}).\\[2pt]
\textsuperscript{d} Particle Data Group (PDG) estimate~\cite{PDG2026}.\\[2pt]
\textsuperscript{e} Residue phase calculated at the pole of the relativistic amplitude with the energy-dependent width ($\Gamma_{W,Z}\propto E^2$, $\Gamma_{H}=\mathrm{const.}$) ~\cite{PDG2026}.\\[2pt]
\textsuperscript{f} Breit-Wigner parameters taken from PDG empirical estimates, resulting in strong departure from our unitarity condition.\\[2pt]
\textsuperscript{g} The PDG reports only empirical BW peak parameters. Lacking extracted pole coordinates, $\xi_E$ cannot be calculated, though naively equating $M \approx M_\mathrm{BW}$ yields a strong unitarity violation.\\[2pt]
\textsuperscript{h} Unitary Prediction: the empirical $M_\mathrm{BW}$ is discarded, and the elastic symmetry condition ($\beta \equiv \alpha$) is enforced. The model predicts the true residue phase (as $2\alpha$).\\[2pt]
\textsuperscript{i} S-matrix pole parameters ($\widetilde{M}, \widetilde{\Gamma}$) theoretically estimated from the empirical peak ($M_\mathrm{BW}$ and $\Gamma_\mathrm{BW} = 20.5(25)$ MeV) by enforcing elastic geometry ($\beta \equiv \alpha$) via Eq.~(\ref{eq:pole_reconstruction}). We treat these parameters on equal footing with those of the $\Delta(1232)$ resonance.
\end{table*}

To quantify deviations from pure elastic behavior, we introduce the dimensionless metric $\xi_E$, measuring the displacement of the empirical peak from its unitary expectation in half-width units:
\begin{equation}
\xi_E = \frac{|M_\mathrm{BW} - \widetilde{M}_\mathrm{BW}|}{\Gamma/2}.\label{eq:ksiE}
\end{equation}

By substituting Eqs.~(\ref{eq:alphabeta}) and (\ref{eq:MBWU}) into this definition, this elegantly reduces to the geometric formula $\xi_E = |\tan \alpha - \tan \beta|$, where $\xi_E \approx 0$ signifies pure elastic behavior.

Noting convention differences across subfields, we adopt the excited nucleon convention, as it is uniformly applicable from nuclear resonances to heavy bosons: $|r|e^{i\theta}=\lim_{E\to E_p}(E_p - E)\,T(E)$. 

For meson resonances, which are traditionally analyzed in the Mandelstam $s$-plane, their reported dispersive residue phases $\phi_g$ must be properly translated to our $E$-plane formalism. This kinematic translation absorbs both the Jacobian energy rotation and the complex centrifugal barrier, yielding the exact relation (for a detailed derivation, see Refs.~\cite{CeciMesons}):
\begin{equation}
\theta = 2\phi_g - 2\arg(E_p) + (2\ell+1)\arg(q(E_p)), \label{eq:convention}
\end{equation}
where $q(E_p)$ is the center-of-mass momentum evaluated at the pole and $\ell$ is the angular momentum quantum number.

\begin{figure}[tbp]
\centering
\begin{tikzpicture}
\begin{axis}[
xmode=log,
ymode=log,
width=8.5cm,
height=7.5cm,
xmin=0.4, xmax=300000,
ymin=0.1, ymax=2500,
xlabel={$(M - E_0)$ / MeV},
ylabel={$\Gamma/2$ / MeV},
grid=both,
grid style={dashed, gray!30},
title={Residue phase prediction diagram ($\bm{\theta=2\alpha}$)},
title style={font=\bfseries, yshift=4ex},
xtick={1, 10, 100, 1000, 10000, 100000},
ytick={0.1, 1, 10, 100, 1000},
axis background/.style={fill=white},
clip=false
]

\begin{scope}
\clip (0.4,0.1) rectangle (300000,2500);

\fill[violet!12] (0.1, 1e-4) -- (1e7, 1e-4) -- (1e7, 8.75e5) -- (0.1, 0.00875) -- cycle;
\fill[blue!8] (0.1, 0.00875) -- (1e7, 8.75e5) -- (1e7, 2.679e6) -- (0.1, 0.02679) -- cycle;
\fill[green!8] (0.1, 0.02679) -- (1e7, 2.679e6) -- (1e7, 1e7) -- (0.1, 0.1) -- cycle;
\fill[yellow!15] (0.1, 0.1) -- (1e7, 1e7) -- (1e7, 3.732e7) -- (0.1, 0.3732) -- cycle;
\fill[orange!15] (0.1, 0.3732) -- (1e7, 3.732e7) -- (1e7, 1.143e8) -- (0.1, 1.143) -- cycle;
\fill[red!8] (0.1, 1.143) -- (1e7, 1.143e8) -- (1e7, 1e10) -- (0.1, 1e10) -- cycle;

\addplot [domain=0.1:1000000, gray!80, dashed, thick] {0.0875 * x};
\addplot [domain=0.1:1000000, gray!80, dashed, thick] {0.2679 * x};
\addplot [domain=0.1:1000000, gray!80, dashed, thick] {1.0 * x};
\addplot [domain=0.1:1000000, gray!80, dashed, thick] {3.732 * x};
\addplot [domain=0.1:1000000, gray!80, dashed, thick] {11.43 * x};
\end{scope}

\node[above, font=\small, text=black!80, yshift=3pt] at (15, 2500) {$\bm{\alpha} \rightarrow$};

\node[above, font=\small, text=black!80, yshift=2pt] at (28571, 2500) {$-5^\circ$};
\node[above, font=\small, text=black!80, yshift=2pt] at (9331, 2500) {$-15^\circ$};
\node[above, font=\small, text=black!80, yshift=2pt] at (2500, 2500) {$-45^\circ$};
\node[above, font=\small, text=black!80, yshift=2pt] at (670, 2500) {$-75^\circ$};
\node[above, font=\small, text=black!80, yshift=2pt, xshift=2pt] at (160, 2500) {$-85^\circ$};

\draw[black!60, thin] (28571, 2500) -- ++(0pt, 3pt);
\draw[black!60, thin] (9331, 2500) -- ++(0pt, 3pt);
\draw[black!60, thin] (2500, 2500) -- ++(0pt, 3pt);
\draw[black!60, thin] (670, 2500) -- ++(0pt, 3pt);
\draw[black!60, thin] (219, 2500) -- ++(0pt, 3pt);


\addplot[only marks, mark=triangle*, mark options={draw=black, thick, fill=violet, scale=2.2, rotate=180}] coordinates {
(80340.5, 1070) 
(91161, 1247)   
(125250, 1.6)   
};
\node[left, xshift=-4pt] at (80340.5, 1070) {$W$};
\node[right, xshift=4pt] at (91161, 1247) {$Z$};
\node[below, yshift=-4pt] at (125250, 1.6) {$H$};

\addplot[only marks, mark=*, mark options={draw=black, thick, fill=blue, scale=1.8}] coordinates {
(483.5, 73.2) 
(259, 29) 
};
\node[right, xshift=4pt, yshift=3pt] at (483.5, 73.2) {$\rho(770)$};
\node[right, xshift=4pt] at (259, 29) {$K^*(892)$};

\addplot[only marks, mark=diamond*, mark options={draw=black, thick, fill=green, scale=2.6}] coordinates {
(132, 50) 
};
\node[left, xshift=1pt, yshift=4pt] at (132, 50) {$\Delta(1232)$};

\addplot[only marks, mark=pentagon*, mark options={draw=black, thick, fill=yellow!90!black, scale=2.0}] coordinates {
(178, 279) 
};
\node[right, xshift=4pt, yshift=6pt] at (178, 279) {$f_0(500)$};

\addplot[only marks, mark=square*, mark options={draw=black, thick, fill=red, scale=1.8}] coordinates {
(15, 280) 
};
\node[left, xshift=-4pt, yshift=3pt] at (15, 280) {$K_0^*(700)$};


\addplot[only marks, mark=triangle*, mark options={draw=black, thick, fill=white, scale=2.2, rotate=180}] coordinates {
(70.3, 4.5)  
};
\node[right, xshift=2pt, yshift=-6pt] at (70.3, 4.5) {$\Xi(1530)$};

\addplot[only marks, mark=*, mark options={draw=black, thick, fill=white, scale=1.8}] coordinates {
(124, 17.5)  
};
\node[right, xshift=2pt, yshift=-6pt] at (124, 17.5) {$\Sigma(1385)$};

\addplot[only marks, mark=diamond*, mark options={draw=black, thick, fill=white, scale=2.6}] coordinates {
(0.8, 0.325)  
(17.4, 8.34)  
};
\node[right, xshift=1pt, yshift=4pt] at (0.8, 0.325) {$^5\text{He}$};
\node[left, xshift=-1pt, yshift=5pt] at (17.4, 8.34) {$\Upsilon(4S)$};

\end{axis}
\end{tikzpicture}
\caption{The $\alpha$-diagram and our residue phase predictions. Symbol shapes and colored fills encode the independently extracted dispersive residue phase ($\theta_{Ref}$), exactly as in Fig.~\ref{fig:apparent_chaos}. Here, they fit their respective background bands (Breit-Wigner region for $\alpha$ from $-5^\circ$ to $0^\circ$, typical resonances region from $-45^\circ$ to $-5^\circ$, and non-Breit-Wigner region with $\alpha$ less than $-45^\circ$), confirming unitarity constraint $\alpha=\beta$ and H\"ohler's $\theta= 2\beta$. States that had no residue phase estimates (designated as $\times$ or a question mark), now sort themselves into distinct geometric sectors. They are plotted using empty (unfilled) symbols, inviting future empirical validation. For the $\Upsilon(4S)$ state, not only its missing residue phase, but also its pole ($\widetilde{M}, \widetilde{\Gamma}$) are estimated from the BW parameters of its measured peak.}
\label{fig:kinematic_landscape}
\end{figure}
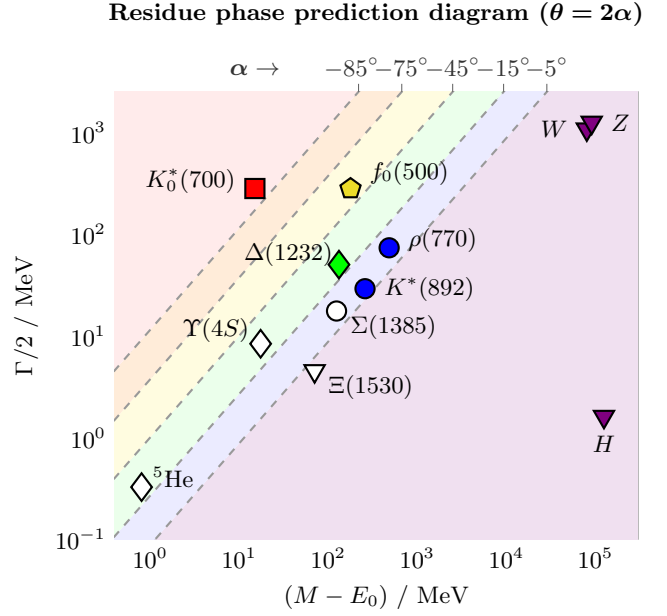

\textit{Results.}---To test the implications of imposing elastic unitarity, we systematically apply our framework to resonances across a broad range of energies. The results are summarized in Table~\ref{tab:main_results} and visualized in Fig.~\ref{fig:kinematic_landscape}.

Part I of Table~\ref{tab:main_results} shows canonical elastic hadrons and nuclear $^5\text{He}$ obeying $\xi_E \approx 0$, confirming $\beta \approx \alpha$. Where empirical phases exist, our geometric predictions ($\theta = 2\beta$) prove highly accurate.

In Part II, we apply our framework to heavy elementary particles. Despite their multichannel decays, this is justified because their full $T$-matrix factorizes into a constant branching-fraction matrix multiplied by a unitary elastic-like scalar amplitude~\cite{PDG2026}, which exclusively defines the residue phase. For the $W$ and $Z$ bosons, this scalar amplitude features an energy-dependent width. Calculating the exact complex poles directly from these standard amplitudes yields reference phases that match H\"ohler's rule ($\theta = 2\beta$). Due to their vast distance from the threshold, the geometric equivalence ($\beta \approx \alpha$) becomes cruder, marking the natural limit of our $\theta \approx 2\alpha$ prediction. Conversely, the Higgs boson has a constant width. Using the same approach as for the $W$ and $Z$ bosons yields no shift between its $M$ and $M_\mathrm{BW}$, and a zero residue phase. A very distant threshold is consistent with $\alpha=0^\circ$. The elastic relation $\beta=\alpha$ predicts exactly the same thing: no mass shift and a zero phase. The Higgs boson is a perfect example of the classical Breit-Wigner resonance.

Part IIIa reveals a strong departure from the elastic unitarity constraint ($\xi_E \gg 0$) for broad scalars $f_0(500)$ and $K_0^*(700)$ when using PDG BW masses. To resolve this discrepancy in Part IIIb, we discard these empirical peak (BW) masses entirely, especially since classical peaks cannot be seen anywhere near the pole mass $M$ in the corresponding phase-shift analyses~\cite{PR,GM}. Instead, we rely on the threshold and pole to determine $\alpha$, and enforce our unitarity condition ($\beta \equiv \alpha$, setting $\xi_E \equiv 0$).

This yields phase predictions ($-115(5)^\circ$ for $f_0$ and $-174(3)^\circ$ for $K_0^*$), in exceptional agreement with sophisticated dispersive complex pole (CP) extractions~\cite{GM, PR, CeciMesons}, which yield $\theta_{CP} = -118(12)^\circ$ and $-180(8)^\circ$, respectively. We note that for the $\rho(770)$ and $f_0(500)$ mesons, these geometric predictions are also consistent with the recent high-precision Roy-equation extractions from Ref.~\cite{HR}, yielding $\theta = -19.0(6)^\circ$ and $-124(10)^\circ$, respectively.

Finally, having validated our geometric constraint on the broad scalars, we apply the inverse procedure (Eq.~\ref{eq:pole_reconstruction}) to the $\Upsilon(4S)$. Using only its empirical BW parameters ($M_\mathrm{BW} = 10579(1)$ MeV and $\Gamma_\mathrm{BW} = 20.5(25)$ MeV)~\cite{PDG2026}, our analytical reconstruction defines its S-matrix pole at $\widetilde{M} = 10575(1)$ MeV and $\widetilde{\Gamma} = 16.6(13)$ MeV, predicting its residue phase to be $-52(6)^\circ$.

\textit{Discussion.}---To clarify our geometric relation further, we write the elastic K-matrix (reactance) with a resonance pole at the empirical BW mass and an energy-dependent width:
\begin{equation}
K(E) = \frac{\Gamma(E)/2}{M_\mathrm{BW} - E}. \label{eq:kmatrix}
\end{equation}
Following Manley~\cite{Manley}, we assume the width varies slowly near the peak $\Gamma(E) \approx \Gamma_\mathrm{BW} + 2(M_\mathrm{BW} - E)\tan\beta$. Using $T = K/(1-iK)$ to reconstruct the scattering amplitude yields an S-matrix pole that reproduces Eq.~(\ref{eq:pole_reconstruction}). Evaluating the S-matrix residue at this pole yields its modulus $|r| = \widetilde{\Gamma}/2$ (where $\widetilde{\Gamma}$ is the pole width $\Gamma_\mathrm{BW}\,\cos^2\beta$) and the phase $\theta = 2\beta$. Our key contribution here is recognizing that the energy-dependent width vanishes at the reaction threshold, and thus this form must have an effective zero there. Imposing this condition ($K(E_0)=0$) on our linear approximation reproduces the geometric relation $\beta=\alpha$ in Eq.~(\ref{eq:alphaBW}).

In order to test our pole reconstruction formula, we choose the $\Delta(1232)$ resonance since it has well-established pole and BW parameters. We obtain $\widetilde{M} = 1213(2)$ MeV and $\widetilde{\Gamma} = 102(3)$ MeV, which are within 2--3 MeV of PDG estimates. Furthermore, the residue phase, calculated as $2\alpha$, yields a reasonable estimate of $\theta = -42(1)^\circ$. This justifies our $\Upsilon(4S)$ extraction.

\textit{Conclusion.}---We have demonstrated that the relationship between the S-matrix pole, the scattering threshold, and the observable peak of elastic resonances is governed by unitarity. The angle between the pole and the real energy axis as seen from the threshold ($\alpha$) classifies states into three distinct geometric regions. For $\alpha$ between $0^\circ$ and $-5^\circ$, states fall into the Breit-Wigner region (exemplified here by the Higgs boson), where the original 1936 BW formula holds and the residue phase is practically zero. The $-5^\circ$ to $-45^\circ$ range defines typical resonances, exhibiting a noticeable peak shift from the pole mass $M$ and a distinct, but relatively small, residue phase. Finally, the $-45^\circ$ to $-90^\circ$ sector hosts non-BW resonances, such as the $f_0(500)$ and $K_0^*(700)$ mesons, where the peak is significantly shifted or entirely absent from the vicinity of the pole, accompanied by huge negative residue phases. By inverting our formalism, we have provided a way to estimate elastic resonances' pole parameters from empirical BW properties.

\end{document}